\documentclass{sig-alternate}
\newfont{\mycrnotice}{ptmr8t at 7pt}
\newfont{\myconfname}{ptmri8t at 7pt}
\let\crnotice\mycrnotice%
\let\confname\myconfname%

\usepackage[german,american]{babel}
\usepackage[T1]{fontenc}
\usepackage[utf8]{inputenc}
\usepackage{times}
\usepackage{textcomp}
\usepackage{graphicx}
\usepackage{amssymb}
\usepackage{microtype} 

\usepackage{url}
\renewcommand{\tt}{\fontencoding{OT1}\fontfamily{cmtt}\selectfont}
\newcommand{\RNum}[1]{\uppercase\expandafter{\romannumeral #1\relax}}

\usepackage[usenames,dvipsnames]{color}
\usepackage{booktabs}
\usepackage[numbers,sectionbib]{natbib}
\makeatletter
\def\@copyrightspace{\relax}
\makeatother

\usepackage{footnote}

\begin{document}

\title{Inequalities, Preferences and Rankings in  US Sports  Coach Hiring Networks}

\numberofauthors{3}
\author{
\alignauthor
Huanshen Wei\footnotemark[1]\footnotetext[1]{This work was largely conducted while the author visited Penn State as an  intern in the Summer of 2017.}\\ \vspace{0.1in}
\affaddr{School of Electronics and Information Engineering\\ Beihang 
University, China}\\
\vspace{0.05in}
\affaddr{huanshen.wei@ist.psu.edu}\\
\alignauthor
Jason (Jiasheng) Zhang\\ \vspace{0.1in}
\affaddr{College of IST\\Penn State University, USA}\\
\vspace{0.05in}
\affaddr{jpz5181@ist.psu.edu}\\
\and
\alignauthor
Dongwon Lee\\ \vspace{0.1in}
\affaddr{College of IST\\Penn State University, USA}\\
\vspace{0.05in}
\affaddr{dlee@ist.psu.edu}\\
}
\maketitle
\renewcommand{\thefootnote}{\fnsymbol{footnote}}
\footnotetext[1]{This work was largely conducted 
while the author visited Penn State as an  intern in the Summer of 
2017.}

\begin{abstract}
Hiring a head coach of a college sports team is vital which will definitely have a great influence on the later development of the team. However, a lot of attention has been focused on each coach's individual features. A systematic and quantitative analysis of the whole coach hiring market is lacking. In a coach hiring network, the coaches are actually voting with their feet. It is interesting to analyze what factors are affecting the "footprint" left by those head coaches. In this paper, we collect more than 12,000 head coach hiring records in two different popular sports from the NCAA. Using network-based methods, we build the coach hiring network in the NCAA men's basketball and football. We find that: (1).the coach hiring network is of great inequality in coach production with a Gini coefficient close to 0.60. (2).coaches prefer to work within the same geographical region and the same division to their alma maters'. (3).the coach production rankings  we calculated using network-based methods are generally correlated to the authoritative rankings, but also show disaccord in specific time period. The results provide us a novel view and better understanding of the coach hiring market in the NCAA and shed new light on the coach hiring system.
\end{abstract}

\section{Introduction}
College sports organized The National Collegiate Athletic Association(NCAA) are very popular in the United States. And the most popular sports include football and men's basketball. According to a survey conducted by Harris Interactive, NCAA's inter-college competition is attracting around 47 percent of the U.S. Americans. In 2016, about 31 million of the followers have attended a college sports event. In the meanwhile, the enormous number of the followers indicates a big business behind the sport events which generated 740 million U.S. dollars in revenue from television and marketing rights in 2016.

The fans and media have paid a lot of attention to the head coaches' salaries and the coaching changing each year. However, the extent of the analysis is limited to unofficial news comments and folk discussion. A quantitative analysis of the head coaches' hiring is lacking. Most of the head coaches used to be excellent players in his/her college team. Therefore, the defense/offense technique that he/she learnt and was familiar with during the college serving time will have an important impact on his/her later coach career. When a school $u$ hires a graduate from school $v$ as its head coach, $u$ implicitly makes a positive assessment of the quality of $v$’s sports program. By collecting this kind of pairwise assessment, we built the coach hiring networks. we use network-based methods to analyze the head coach hiring networks during the years.

Contributions of our work are as follows: 1.We find high inequality in the coach hiring networks, which means that most of the head coaches graduated from a small proportion of the schools. 2.Based on optimal modularity, We find geographic communities of the schools. A graduate from one community is more likely to be hired as a head coach of a school in the same community. 3.Our coach production rankings have shown general correlation to the authoritative Associated Press(AP) rankings while some disparity do exist. 4. We find a common within-division flows pattern from the division-level movements of the coach.

\section{Background}
A. Clauset, S. Arbesman, and D.B Larremore’s research article Systematic inequality and hierarchy in faculty hiring networks~\cite{clauset2015systematic} analyzed the academic faculty hiring networks across three disciplines in computer science, history, and business. We are curious about if this kind of inequality and hierarchy also exist in the sports coach hiring network. Fast, Andrew and Jensen, David's work~\cite{fast2006nfl} use the NFL coaching network to identify notable coaches and to learn a model of which teams will make the playoffs in a given year. To identify notable coaches, their networks focused on the work-under relationship between the coaches. Although there have been some papers researched on ranking the sports teams based on their game results~\cite{park2005network,callaghan2007random}, none of them utilized the coach hiring network which is actually an assessment network built of those professional sports experts' view.  

\section{Data Description}
From the official site of the NCAA~\cite{CoachData}, we collect a list of the head coaches(including those retired ones) of the NCAA men's basketball and football teams as well as their coaching career data which includes the coaches' alma maters, graduation year and the school they worked for during their head coaching career. Here we only take the head coaches into account for other positions' data, like the assistant coach, are mostly incomplete which means that the alma maters, the graduation year are difficult to identify. At the same time, we also have removed the head coaches with missing alma mater and graduation time. 

\begin{table}[h]
\centering
\begin{tabular}{|p{2.3cm}|p{1.4cm}|p{1.4cm}|}
\hline
& Football & Basketball \\ 
\hline
schools & 857 & 1214\\
\hline
head coach & 5744 & 6906\\
\hline
mean degree & 6.70 & 5.69\\
\hline
self-loops & 18.35\% & 18.98\%\\
\hline
mean hiring years & 6.33 & 7.03\\
\hline
data period & 1880-2012 & 1888-2013\\
\hline
\end{tabular}
\caption{Coach hiring networks data summary} 
\label{table:NetData}
\end{table}

For each school $u$ that a coach has worked for, a directed edge was generated from the coach's alma mater $v$ to the school $u$. We then extract those schools connected by the edges. A brief networks data summary is listed as Table~\ref{table:NetData}. Almost one-fifth of the head coaches would finally have a chance to serve at their alma maters. This result is much higher than the one in faculty hiring network~\cite{clauset2015systematic}.

Besides, we collect the division attribute of the schools~\cite{DivData}, the authoritative ranking: the Associated Press(AP) rankings data of the two sports~\cite{APRankingData}.

\section{Experimental Methods and results}

\subsection{Head Coach Production Inequality}
We measure the inequality of the coach hiring networks since the two sports were introduced to colleges. Table~\ref{table:InequlData} summaries the basic inequality measurements of our experiment. 

\begin{table}[h]
\centering
\begin{tabular}{|p{2.3cm}|p{1.4cm}|p{1.4cm}|}
\hline
  & Football & Basketball \\ 
\hline
vertices & 857 & 1214\\
\hline
edges & 5744 & 6906\\
\hline
50\% coach from & 14.24\% & 15.16\%\\
\hline
Gini, $G(k_o)$ & 0.59 & 0.58\\
\hline
Gini, $G(k_i)$ & 0.39 & 0.35\\
\hline
$k_o/k_i>1$ & 33.96\%(291) & 33.20\%(403)\\
\hline
\end{tabular}
\caption{measures of inequality in coach hiring networks: percentage of schools required to cover 50\% of the head coaches; Gini coefficient of production(out-degree) and hiring(in-degree); percentage of schools produced more coaches than the number of the coaches it hired} 
\label{table:InequlData}
\end{table}

\begin{figure}[htbp]
	\centering
 	\includegraphics[height = 7cm, width = 8cm]{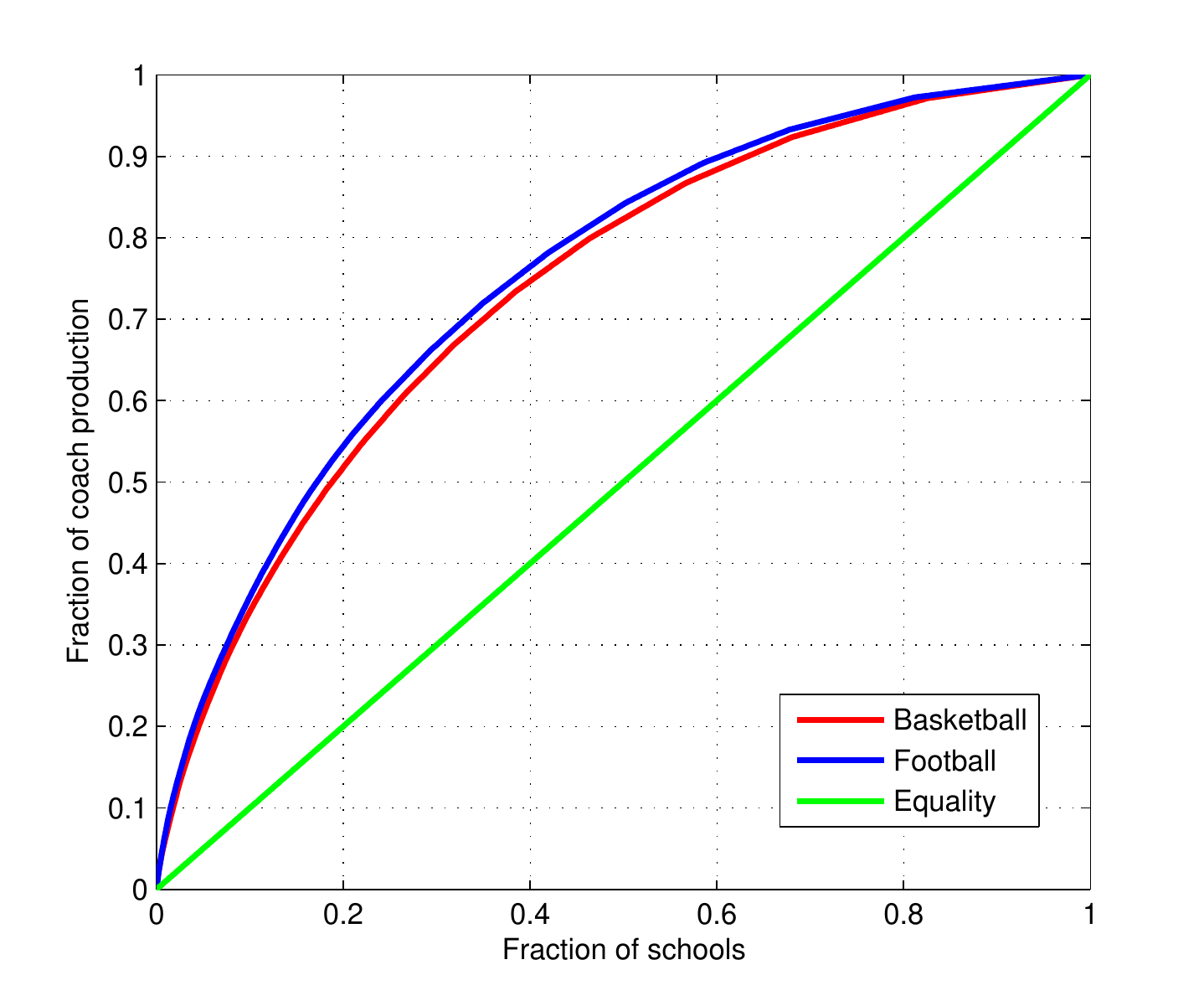}
 	\caption{Coach Production Lorenz Curve}
	\label{fig:Lorenz}  
\end{figure}

The Gini coefficient is the most commonly used measure of inequality. A Gini coefficient of zero means a perfect equality. On the opposite, a maximal inequality will lead to a Gini coefficient of one. Here we respectively calculate the Gini coefficient of the coach production(out-degree) and the coach "consumption"(in-degree). We find that the Gini coefficients of coach production are close to 0.60 which indicate a strong inequality. The income distribution Gini index of South Africa estimated by World Bank in 2011 is 0.63. Figure~\ref{fig:Lorenz} is the Lorenz Curve of the coach production. From the curve, to cover 50\% of the head coaches, it only need around 15\% of the schools, which means that a small proportion of the schools have produced a lot of head coaches to all the NCAA members.

\subsection{Community structure of Coach Hiring Networks}
Community structure is an important property of complex networks. Here we use the modularity optimization algorithm~\citep{blondel2008fast} to detect communities in the coach hiring networks.

Both in the football and men's basketball coach hiring networks, the average modularity of the whole networks are beyond 0.40, which indicates a significant community structure~\cite{newman2004fast}. Both the networks consist of 6 big communities which  includes 97.8\% of schools in the football network and 98.6\% of schools in men's basketball network.

\begin{figure*}[htbp]
	\centering
 	\includegraphics[height = 9.6cm, width = 16.8cm]{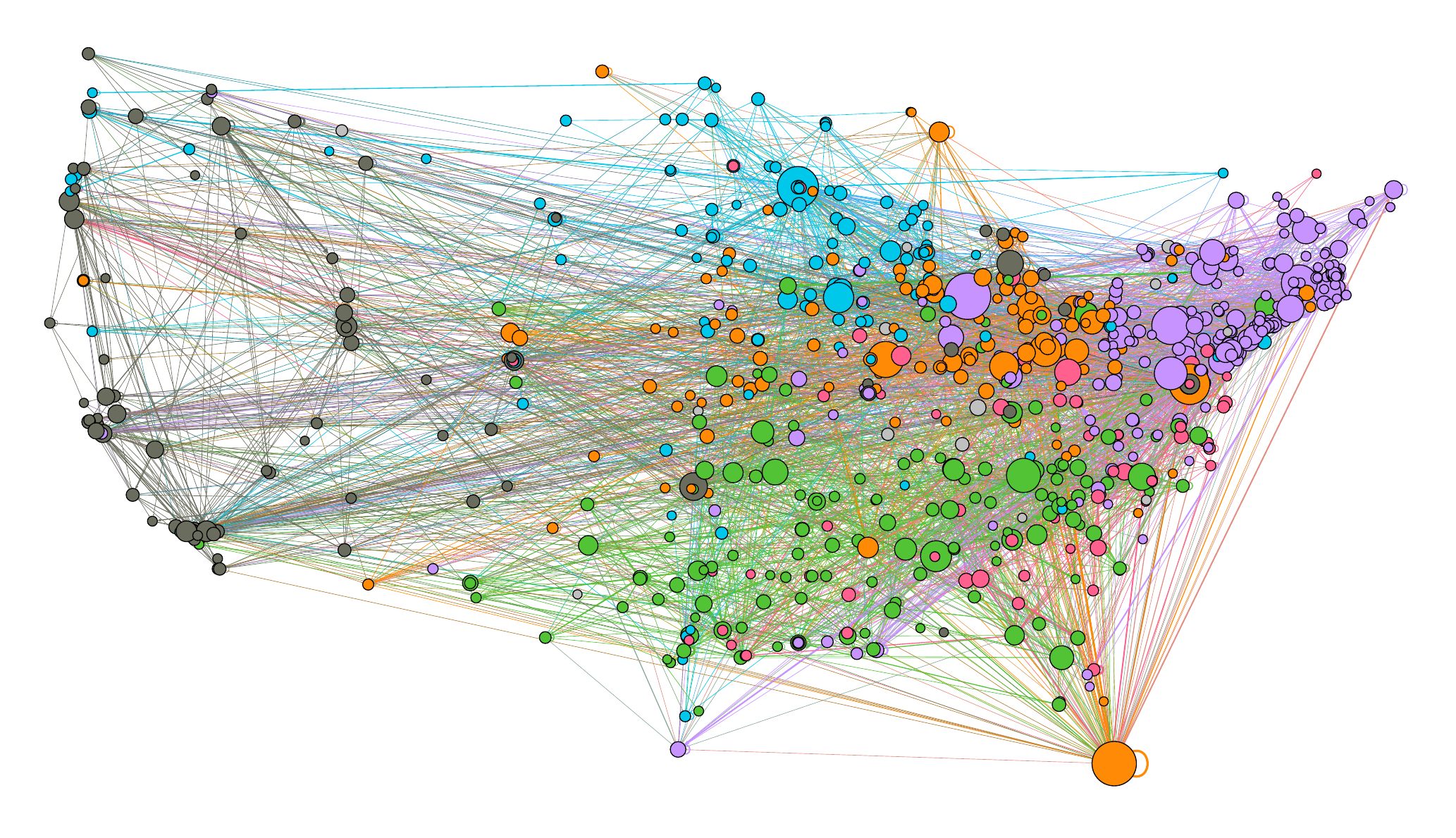}
 	\caption{American Football Coach Hiring Network}
	\label{fig:FootballGeoGraph}  
\end{figure*}

\begin{figure*}[htbp]
	\centering
 	\includegraphics[height = 9.6cm, width = 16.8cm]{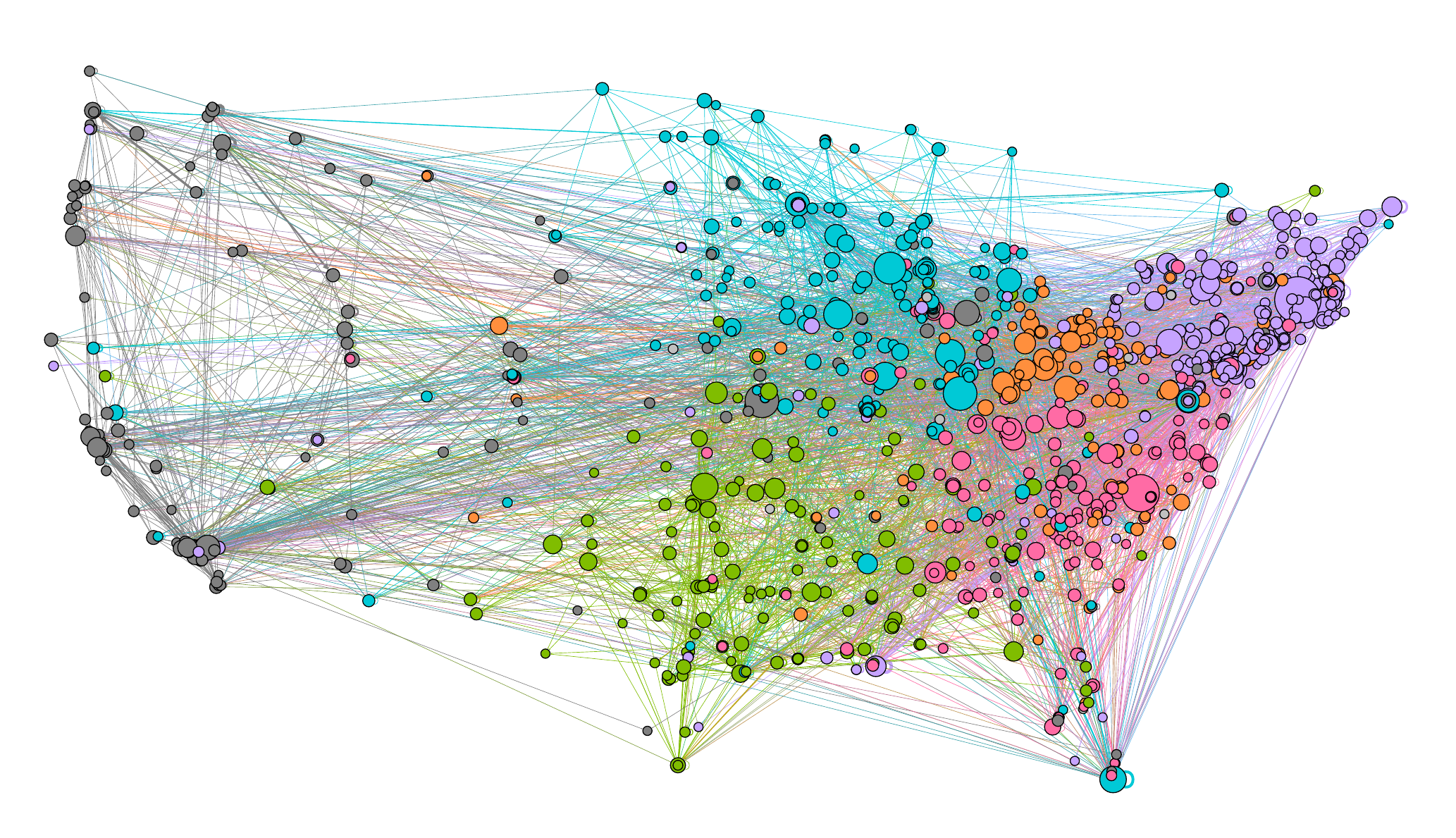}
 	\caption{American Men's Basketball Coach Hiring Network}
	\label{fig:BasketballGeoGraph}  
\end{figure*}

We make a visualization of the networks as Figure~\ref{fig:FootballGeoGraph} and Figure~\ref{fig:BasketballGeoGraph} in which we place each school according to its longitude and latitude and set the size of node proportional to its coach production(out-degree). In the figures, the top 6 biggest communities are assigned several specific colors(To better present the figures in a proper size, we remove several schools which locate in Hawaii, Alaska and Puerto Rico). From the figures, we find that the community distribution indeed is influenced by the geographic factor. Besides, both the biggest communities in the two networks locate in the northeast part of America(in purple).

\subsection{Correlation with the Authoritative Rankings}
\subsubsection{Temporal characteristic of coach hiring networks}
\begin{figure}[htb]
	\centering
 	\includegraphics[height = 6cm, width =8.6cm]{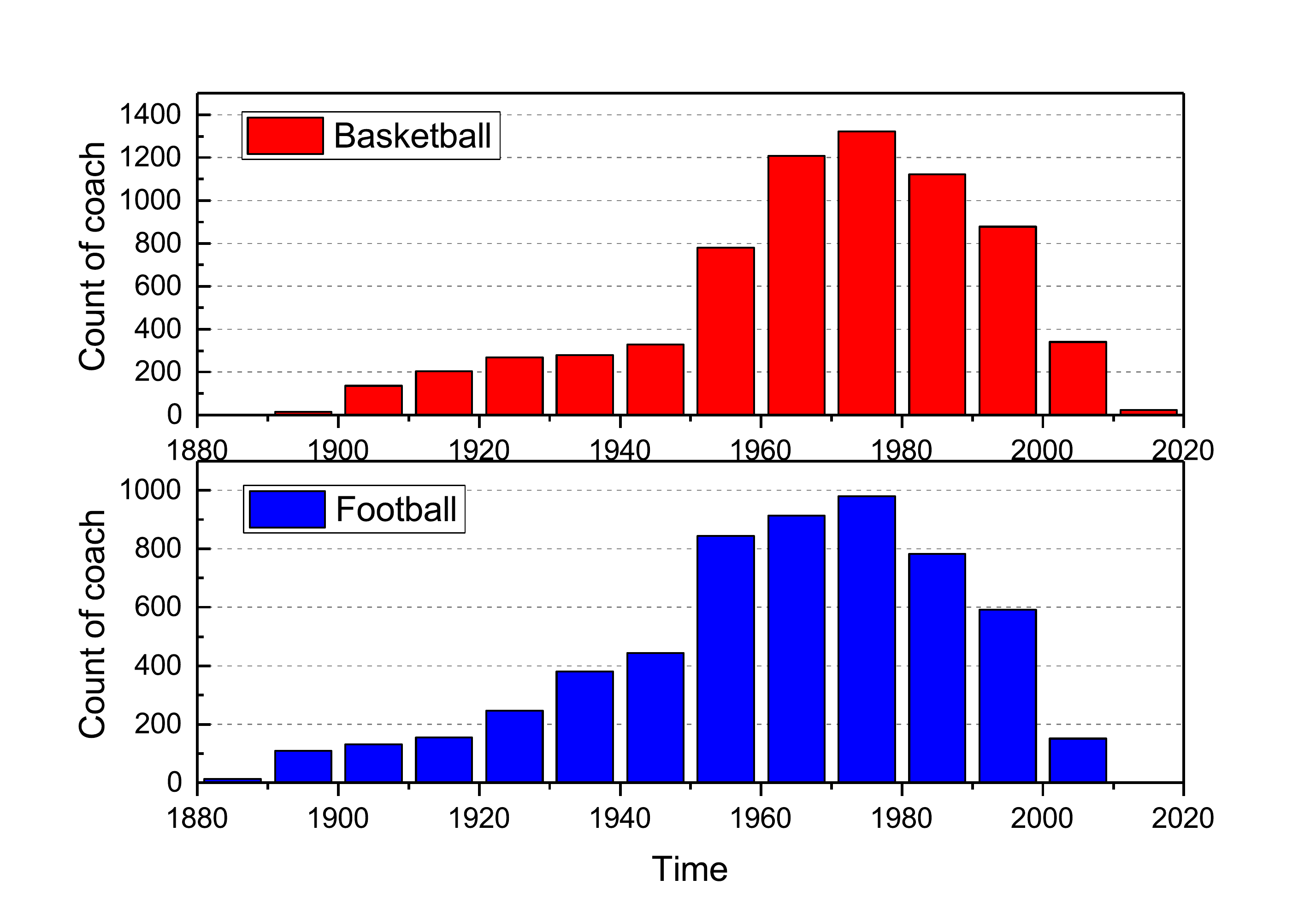}
 	\caption{Counts of Coach in Graduating Year}
	\label{fig:GraDensity}  
\end{figure}

Our data set includes the head coach hiring data roughly from 1980 to 2010. Figure~\ref{fig:GraDensity} is the counts of the coach according to their graduating year. The distribution of the coach records in each period is not even. In fact, the national headquarters of the NCAA was established in Kansas City, Missouri in 1952. So it is not surprised that there are much more head coach records after 1950. 

\begin{figure}[htb]
	\centering
 	\includegraphics[height = 6cm, width =8.6cm]{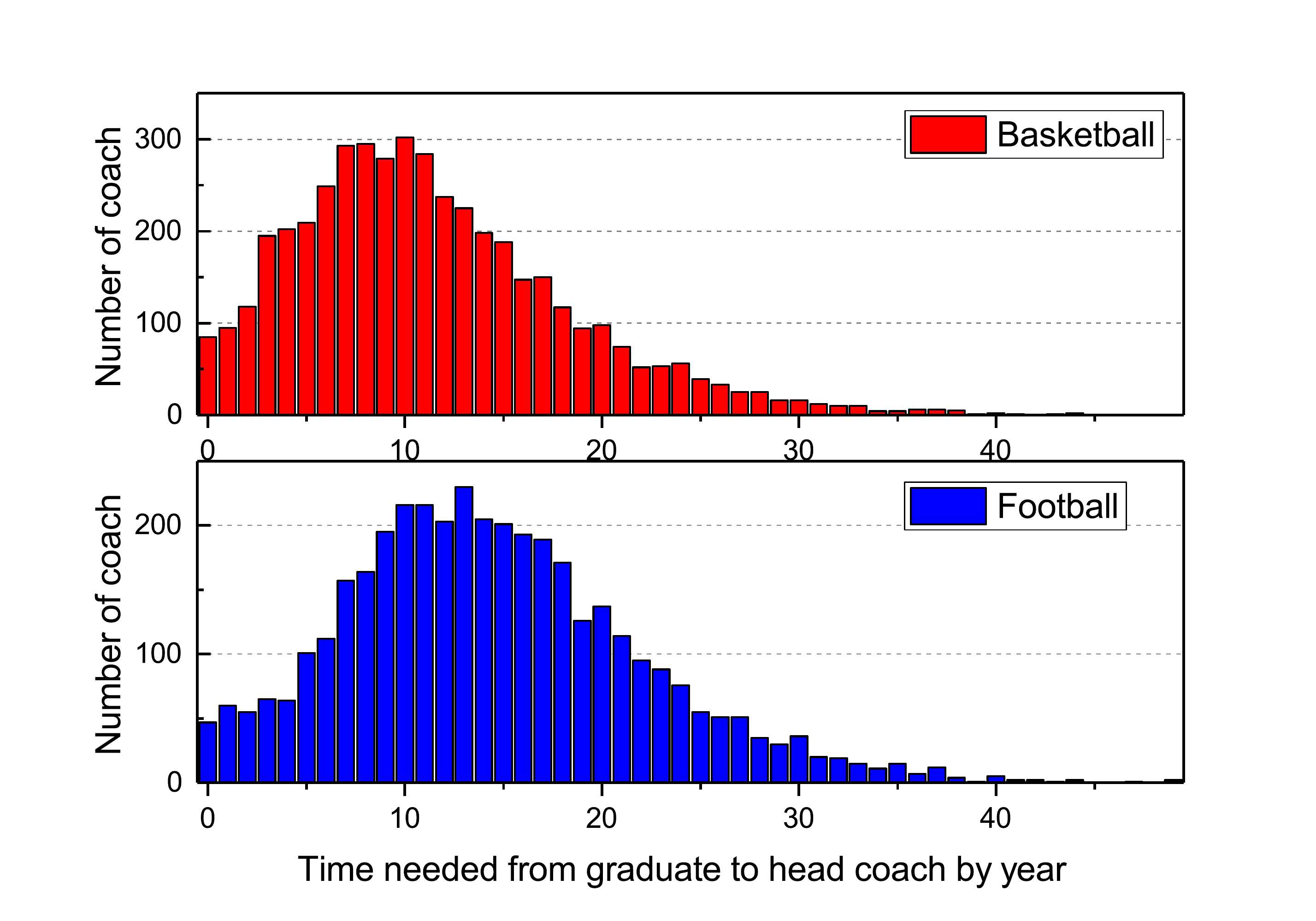}
 	\caption{Histogram of the time needed to become head coach}
	\label{fig:GrowTime}  
\end{figure}

Figure~\ref{fig:GrowTime} shows the counts of coach by the "growing" time needed from graduating to head coach position. The average time needed is 11.5 years in basketball and 14.6 years in football, which means that it takes more time for a graduate to become a football head coach than a basketball head coach. What's more, it also illustrate the scarceness of the head coaches who graduate after 2000 for most of the potential head coach  haven't grown into head coach.

\subsubsection{Coach Production Ranking and Authoritative Rankings}
We want to find out whether the strong teams will foster potential future head coaches. So we generate the coach production rankings from our dataset and calculate the correlation coefficient between the rankings and the authoritative rankings along the years.

Due to the temporal characteristic of the coach hiring records, we extract subnetworks from the whole coach hiring network which only include the coaches graduated in the period as interval $[t_s,t_e]$. Because the number of records in each year is not evenly distributed. We enumerate the $t_e$ from the latest graduating time to the older ones and calculate the corresponding $t_s$ of each interval, ensuring that each interval contains only 30\% of the coaches. Finally, we have 55 subnetworks for men's basketball and 62 subnetworks for football. 

Based on these subnetworks, we try 4 different network-based methods: Out-degree, MVRs~\cite{clauset2015systematic}, PageRank~\cite{brin1998anatomy} and LeaderRank~\cite{lu2011leaders} to rank the schools. The most simple one is based on the out-degree(coach production) of each school. A minimum violation ranking (MVR) is a permutation $\pi$ that induces a minimum number of edges that point “up” the ranking~\cite{clauset2015systematic}. This method try to produce a ranking to minimize the number of coaches who go downward the hierarchy ranking of the schoos. Besides, PageRank is a widely used node ranking method. To apply this algorithm on our dataset, we make the direction of the edges backward. The backward directed edges represent the "votes" from the schools to decide which school's graduates are more welcome. And more importantly, the algorithm also takes into account some schools which produce very few coaches to those powerful schools with a high  in-degree. LeaderRank is an improved version of PageRank in recent year. The Pearson correlation coefficients between the four rankings are all above 0.84, which indicates that these rankings are highly similar to each other. Considering that the PageRank could also find out those important schools with a low degree, we simply choose the traditional PageRank(PR) rankings as the coach production rankings.

We choose the authoritative Associated Press(AP) Poll rankings to compare with our coach production rankings for the reason that polls voting system is also based on subjective opinions of experts. In addition, AP Poll is of long history which is suitable for comparing with our temporal dataset. We aggregate the AP rankings in every 20 years using Median rank aggregation~\citep{fagin2003efficient}. Firstly, we build a school list of the teams which have received votes in every 20 years. Then, In a certain year of the 20-year period, an average rank $(m + 1 + n)/2$ is assigned to the schools not received votes(m represents the number of schools which received votes in the certain year). Finally, we aggregated the 20 rankings into a ranking using Median Rank Aggregation.

\subsubsection{Results}
\begin{figure}[htb]
	\centering
 	\includegraphics[height = 6.5cm, width =8.3cm]{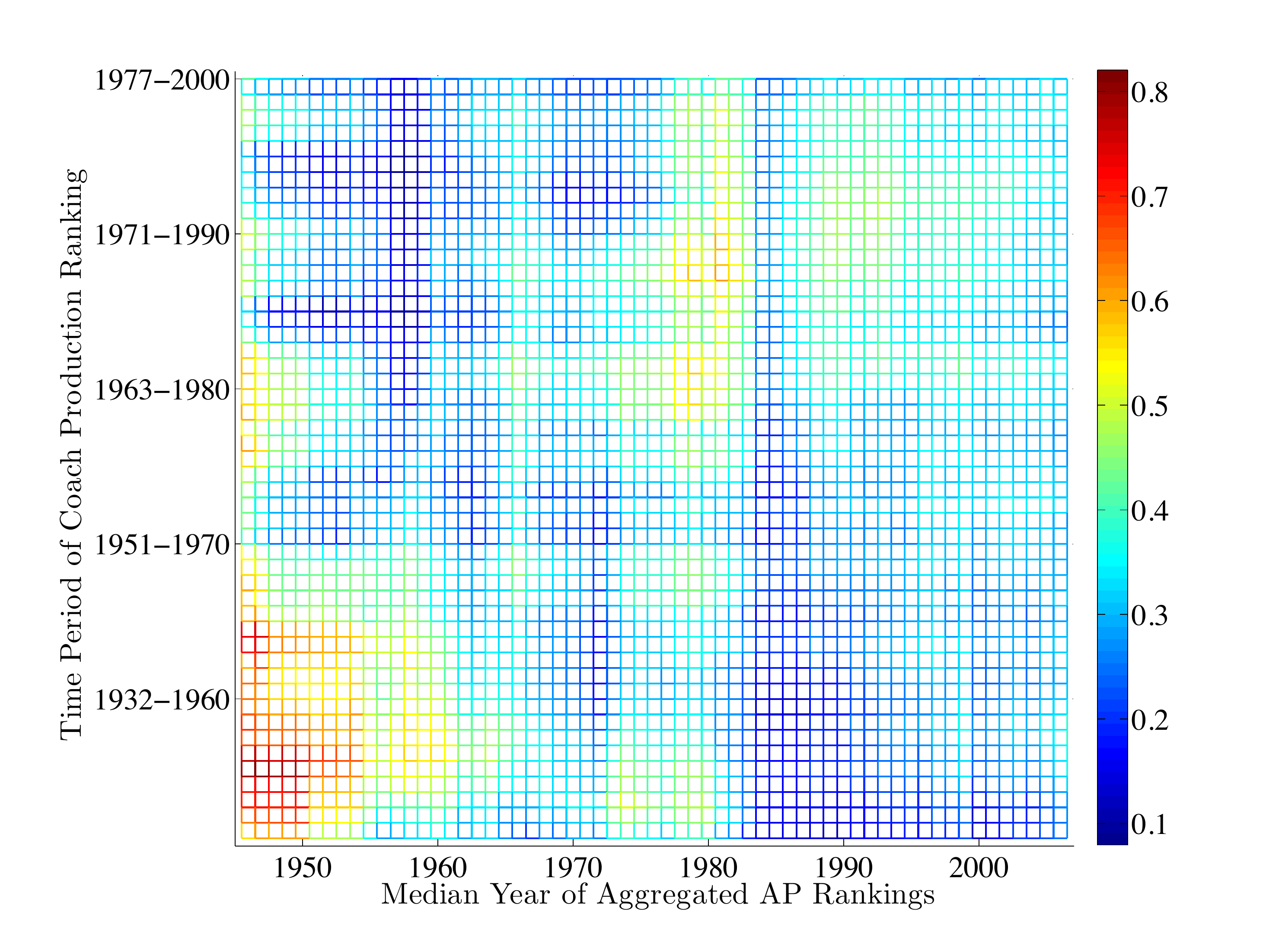}
 	\caption{Correlation Graph Between Aggregated Football AP and PR Rankings}
	\label{fig:FB_Corr}  
\end{figure}

\begin{figure}[htb]
	\centering
 	\includegraphics[height = 6.5cm, width =8.3cm]{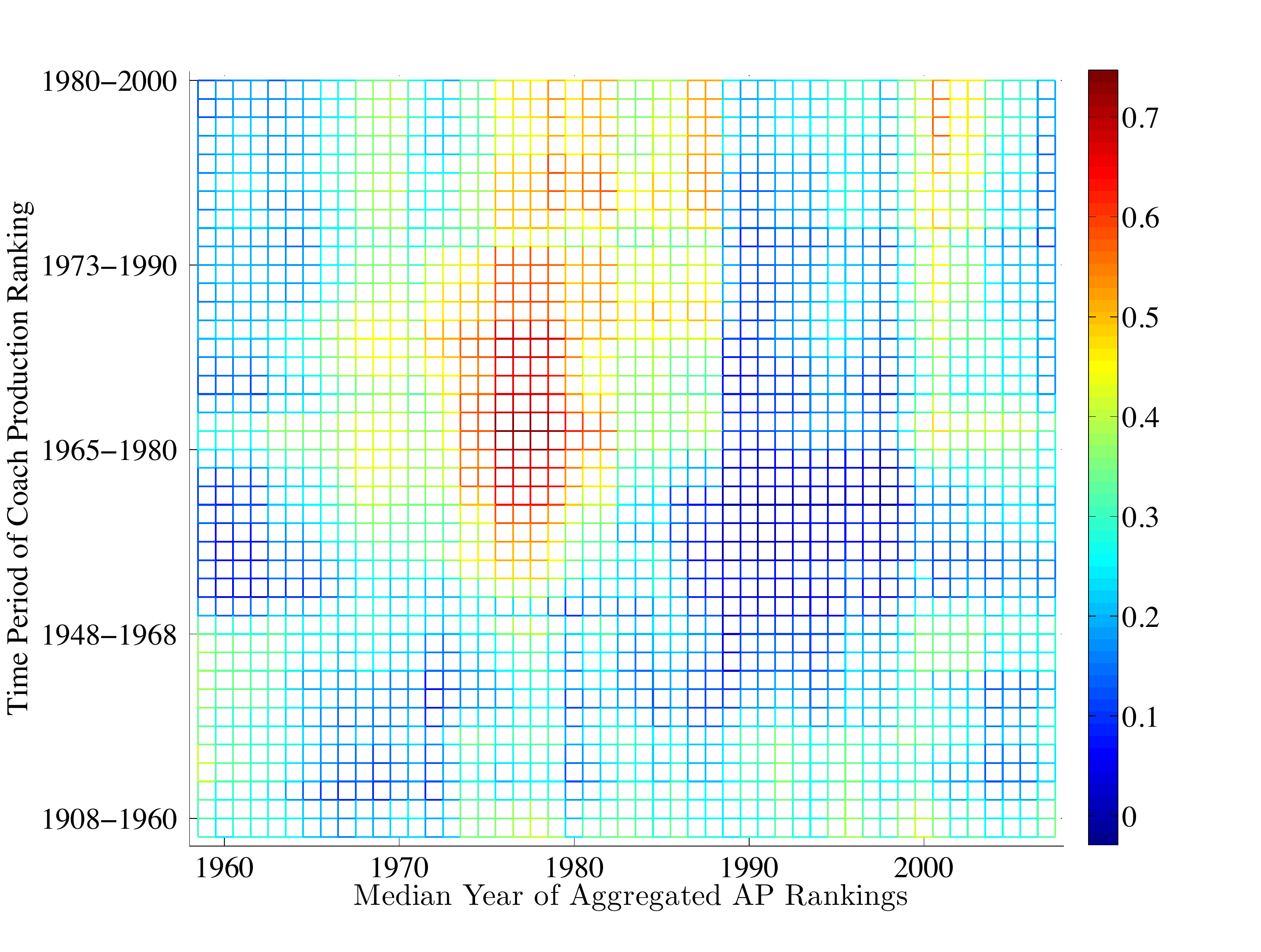}
 	\caption{Correlation Graph Between Aggregated Men's Basketball AP and PR Rankings}
	\label{fig:BK_Corr}  
\end{figure}

We use Kendall's $\tau$~\citep{kendall1938new} to measure the correlation between the coach production rankings and the aggregated AP rankings. The Kendall correlation between two variables will be high when observations have a similar rank. We calculate the Kendall's tau between an aggregated AP ranking $X$ and $Y$---- the corresponding coach production rank of each school in X. Figure~\ref{fig:FB_Corr} and Figure ~\ref{fig:BK_Corr} are the correlation results. The color of the points in the grid represents the value of Kendall's tau between an aggregated AP ranking(as $x$ coordinate) and an coach production ranking(as $y$ coordinate).

We find that: 1.The aggregated AP rankings before 1960 and the ones around 1970-1980 are more correlated to the contemporary coach production rankings(with $\tau>0.5$). 2.In Figure~\ref{fig:FB_Corr}, before 1955, the aggregated AP rankings also show some correlation with the coach production rankings in more recent time. But the correlation gradually decreased by years. Probably because some strong teams in the old days, like Yale and the Ivy League, gradually get insulated from the national spotlight and finally moved down into I-AA(now as Football Championship Subdivision) starting with the 1982 season. 3.In 1973, the NCAA was divided into three legislative and competitive divisions – I, II, and III. And at the same time, both in football and men's basketball, there is an increased correlation between the contemporary aggregated AP rankings and the coach production rankings roughly after 1970.

\subsection{Division-level Movements}
To show the movements between different divisions, we use data of the coach graduated after 1973 when NCAA  was first divided into three divisions. Table~\ref{table:BK_Div} and Table~\ref{table:FB_Div} show the movements from the coach graduating school's division(the row) to the school they work for(the column).
\begin{table}[htb]
\centering
\begin{tabular}{p{1cm}|p{1cm}p{1cm}p{1cm}|p{1cm}}
  & Div I & Div II & Div III & All \\ 
\hline
Div I & \textbf{0.263} & 0.110 & 0.104 & 0.477\\
Div II & 0.062 & \textbf{0.105} & 0.038 & 0.205\\
Div III & 0.062 & 0.044 & \textbf{0.213} & 0.318\\
\hline
All  & 0.386 & 0.259 & 0.355 & \\
\end{tabular}
\caption{In the NCAA men's basketball, faction of coach who graduated from a school in one division(row) and are hired as head coach in a school from another division(column). Movements inside one division are highlighted in bold.} 
\label{table:BK_Div}
\end{table}

\begin{table}[htb]
\centering
\begin{tabular}{p{1cm}|p{0.6cm}p{0.6cm}p{0.8cm}p{0.9cm}|p{0.6cm}}
  & FBS & FCS & Div II & Div III & All \\ 
\hline
FBS & \textbf{0.153} & 0.063 & 0.054 & 0.043 & 0.314\\
FCS & 0.040 & \textbf{0.078} & 0.046 & 0.033 & 0.198\\
Div II & 0.033 & 0.025 & \textbf{0.104} & 0.038 & 0.201\\
Div III & 0.019 & 0.026 & 0.041 & \textbf{0.200} & 0.286 \\
\hline
All & 0.246 & 0.193 & 0.246 & 0.315 & \\
\end{tabular}
\caption{In the NCAA football, faction of coach who graduated from a school in one division(row) and are hired as head coach in a school from another division(column). Movements inside one division are highlighted in bold. Football Division I has two subdivisions I-A and I-AA (renamed the Football Bowl Subdivision(FBS) and the Football Championship Subdivision(FCS) in 2006)} 
\label{table:FB_Div}
\end{table}

Here we also take into account the movements among the two subdivisions of division I in football and other divisions. Generally speaking, the FBS has more funding, more scholarship and better sport facilities than the FCS. From the tables we find that: 1.The diagonal number represents the fraction of coach who graduated and got hired in the same division. And the fraction of within-division movements is greater than others. 2.There are more coaches move downwards from division I to II and III than move upwards. 3.Excluding the coach working within their division, there are more coach moving upwards to division I than moving to division II or III.

\section{Conclusion}
In this paper, we collect a dataset containing the NCAA men's basketball and football head coach's career data and hiring data. Based on the dataset, we build coach hiring networks and use network-based methods to analyze the hiring networks in four aspects including inequality, community structure, coach production rankings and the movements between divisions. The results reveal that: (1).the coach hiring market is actually of great inequality, which means most of the head coaches come from a small proportion of the NCAA members. It indicates an unequal distribution of the US sports education resource. (2).Coaches prefer to stay in the same division and geographic region to their alma maters'. (3).The coach production rankings are generally correlated to the authoritative rankings, which indicates that good teams are likely to foster future head coaches. However, in specific time period, this is not true probably because of the contemporary NCAA policies and social events.

Our future directions include: 1.We have found some similar hierarchical organization properties as in ~\cite{ravasz2003hierarchical} on our dataset. We could develop proper temporal evolving networks model to predict the coach hiring market. 2.To better explain and find out the mechanism behind our findings, such as the inequality, a complete, deeper understanding of the NCAA's history~\cite{smith2000brief}, the contemporary related policy will be probably of help.

\begin{raggedright}
\bibliography{main}
\end{raggedright}
\end{document}